\documentclass[article,8pt,a4paper]{article}
\usepackage[cm]{fullpage}
\usepackage{lmodern}			
\usepackage{lipsum}
\usepackage{listings}
\usepackage{color,hyperref}
\usepackage{graphicx}			
\usepackage[english]{babel}
\usepackage[utf8]{inputenc}   
\usepackage[bottom]{footmisc}
\usepackage{amsmath,amssymb}
\usepackage{abstract}
\renewcommand{\abstractname}{}    

\usepackage{txfonts}
\usepackage[T1]{fontenc}

\usepackage{microtype} 

\begin{document}
	
	
\title{Fractional derivative order determination from harmonic oscillator damping factor.\\
	}
\bigskip

	\author{Luis Felipe Alves da Silva\textsuperscript{1}
	\thanks{Electronic Mail Adress: \href{mailto:}{silvaluiswr@gmail.com}} , 
	Valdiney Rodrigues Pedrozo Júnior \textsuperscript{1}. \\
	João Vítor Batista Ferreira \textsuperscript{1}. \\
	\bigskip
	\large \textsuperscript{1} Institute of Physics, Federal University of Mato Grosso do Sul. \\
	Campo Grande, MS, Brazil}
	
	\maketitle
	
	\renewcommand{\abstractname}{\vspace{-\baselineskip}} 
	\begin{abstract}
		This article analysis differential equations which represents damped and fractional oscillators. First, it is shown that prior to using physical quantities in fractional calculus, it is imperative that they are turned dimensionless. Afterwards, approximated expressions that relate the two equations parameters for the case that the fractional order is close to an integer number are presented. Following, a numerical regression is made using power series expansion,  and, also from fractional calculus, the fact that both equations cannot be equivalent is concluded. In the end, from the numerical regression data, the analytical approximated expressions that relate the two equations' parameters are refined.
	\end{abstract}
	\bigskip

	\twocolumn


	\section{\label{sec:level1}Introduction}
	
Fractional calculus is still a novelty to many Physics undergraduate students, despite excellent articles that make it easier to understand \cite{KHALIL201465}. Many physical phenomena are represented by first ($\alpha=1$) or second ($\alpha=2$) order linear differential equations with constant coefficients. 
Non-integer order differential equations usage, for example $\alpha=1.82$, is studied by the area known as fractional calculus. Initially restricted to the pure mathematics field, fractional differential equations are also being used to study physical phenomena. Diffusion processes, of viscoelasticity and others, can be described by non-integer order differential equations. \cite{DIF-ANOMALA, DebnathAplic, RevModPhys.86.1169}. The computer algebra systems tools (CAS) ease the burden of fractional calculus concept usage.

The use of these tools is changing the way physicists approach the study of natural phenomena. As an example, this article analysis the damped harmonic oscillator case. This phenomena can, apparently, be described either by an integer order differential equation or by a fractional differential equation \cite{Achar2001}. This duplicity in its description leads to conjecture the existence of an analytical equivalence between those two expressions. The proof of this equivalence existence is difficult to obtain by analytical mathematics.

If power series and numerical calculus are used instead of analytical
mathematics, it is possible to come to a conclusion. This alternative procedure is only practical using computational algebra systems to manipulate series with many terms. This article uses a power series with sixty terms to represent a function which is a fractional differential equation solution. The solution achieved implies the two differential equations have no analytical equivalence. 

In this article, the CoCalc programming environment has been used as the computational algebra system. It allows both analytical as numerical manipulation, as well as a good function graphic visualization. CoCalc, previously known as SageMathCloud \cite{sage}, is an on-line computational platform that offers free access under some conditions and there is no need for installation on the user computer. CoCalc is available in many programming languages, Python being the most used and there are many auxiliary documents on the internet about the software.

The concepts involved on this article are shown on section 2, being the fractional derivative and the mass-spring system. The section 3 presents the results obtained in this work, which shows the easiness and practicality in using computational tools to solve physical complicated problems. This section also analyses the differential equations that represents the damped oscillator and the fractional oscillator. First, it is shown that prior to using physical quantities in fractional calculus, it is imperative that they are turned dimensionless. Afterwards, expressions that relate the two equations parameters for the case that the fractional derivative order is close to the integer derivative order are presented. Then, with the use of a power series expansion it is possible to come to a conclusion that the two equations cannot be equivalent. Lastly, using data from numerical regression, the analytical expressions that relate the parameters from both equations are refined. A discussion and the conclusion are made on the section 4 of this paper.

\section{Fractional Calculus and Classical Physical Models}\label{sec:level1.5}
	
On this section, we briefly comment about the fractional calculus and its use on representing a damped mass-spring system.

\subsection{Fractional Derivative}
	
Fractional calculus is still a novelty to many Physics undergraduate students. Good texts can be found in \cite{Historia1,Historia}. The non-integer order derivative history starts in the beginnings of XVII traditional calculus. It was at that time that Gottfried Leibniz speculated about an $\alpha=1/2$ order derivative applied to the function $f(x)=x$. For a more intuitive understanding about integer order calculus operators it is common to use a geometrical visualization and interpretation. In fractional calculus, this is a problem due to the absence of an easy geometrical interpretation for it. However, this challenge should not be considered as insurmountable, like the interpretation of the complex number $i$ exemplifies. The conceiving of a number with the $i^2 =-1$ property is hard, but with time the complex number theory has been developed, today being a fundamental piece in quantum mechanics and electromagnetism studies.
	
\textbf{Caputo fractional derivative.}   
	
Definitions for Caputo fractional derivative have been obtained from the  thesis \cite{Ishteva}. The real order $\alpha\geq0$ derivative of a function $\xi(\tau)$ defined on the interval $0<\tau<\infty$, $\xi$ and $\tau$ being dimensionless variables, is
\begin{equation}
\frac{d^{\alpha}\xi(\tau)}{d\tau^{\alpha}}\equiv J^{n-\alpha} \{\frac{d^{n}\xi}{d\tau^{n}}\}~,\label{eq:DerCaputo}
\end{equation}
\begin{equation}
J^{n-\alpha}\{\frac{d^{n}\xi}{d\tau^{n}}\}\equiv \frac{1}{\Gamma(n-\alpha)}\int_{0}^{\tau}(\tau-\nu)^{(n-\alpha)-1}\frac{d^{n}\xi(\nu)}{d\nu^{n}}d\nu~,\label{eq:DerCaputo2}
\end{equation}
where $n$ is integer, positive and $n-1<\alpha\leq n$. If $\alpha =n$, the integer order calculus is recovered. The usual definition for gamma function 
$\Gamma$ is used \cite{WolframGamma}.

This way, Caputo definition for the $\alpha$ order fractional derivative of a function $\xi(\tau)$ can be understood as the integral transform of an integer order derivative, Eq.\ref{eq:DerCaputo}

Assuming that the integration inverse operation is a derivative, in a similar way, but not identically, to Cauchy integral theorem \footnote{For $J^{\alpha}$ operator order $\alpha>0$.} \cite{WolframTIC} the following notation equivalence can be made $\left(\frac{d}{d\tau}\right)^{n}\equiv J^{-n}$ .
	
Generally, Caputo fractional derivative does not satisfy the commutative property, but satisfies the following relation
\begin{equation}
\frac{d^{\alpha+\beta}\xi(\tau)}{d\tau^{\alpha+\beta}}=\frac{d^{\alpha}}{d\tau^{\alpha}}\left(\frac{d^{\beta}\xi(\tau)}{d\tau^{\beta}}\right)=\frac{d^{m}\xi(\tau)}{d\tau^{m}}~, \label{eq:DerComutativa}
\end{equation}
$m$ being an integer and positive number such that $m=\alpha+\beta$ ($\beta$ real and positive). This property is demonstrated on page $56$ of Ref.\cite{Kai}. This is interesting because the fractional order derivative becomes a natural extension of the integer order derivative and the integer order derivative can be defined as a composition of fractional order derivatives.

On physics, the rate of change is an important concept. For example, velocity is a rate of change of a particle position as a function of time. Mathematically, this corresponds to an order one derivative: $dx/dt$. According to the equations \ref{eq:DerCaputo} and \ref{eq:DerCaputo2}, the integer order rate of change concept can be applied to the fractional derivative: the fractional derivative of $x(t)$, as a function of $t$, is the integer order rate of change followed by an integral transform of the convolution type \cite{WolframConvolution}.

Caputo definition brings two advantages. The first is that it makes a fractional derivative of a constant to always be equals zero, which makes sense to Physics. The second is that deriving $\xi(\tau)$ initially by $\alpha$ and then by $\beta$ the same result is obtained both by this process and by the $\alpha + \beta$ order derivative. It is generally agreed to use Caputo definition for functions that have time variations and known boundary conditions, which is the case for a harmonic oscillator.

Fractional derivative analytical manipulation can become very difficult if the $\xi(\tau)$ function is not elementary. In this article, we show that it is possible to overcome this problem by expanding the $\xi(\tau)$ function in a polynomial and by using computers to treat polynomials with a large number of terms.
	
Some works \cite{DebnathAplic,Hilfer,Achar2001} presents generalized differential equations ($\alpha$ real and positive) that models classical problems while using fractional calculus. Since those phenomena are already described by integer order differential equations, they become important studies that analyze the equivalence between the two approaches, the fractional and the integer order. This article contributes to this analysis.

\subsection{Spring-Mass System}
	
The spring-mass oscillator system is one of the most simple models in classical mechanics. Beyond being simple, it is very useful to approximately represent more complex physical systems. For example, it is used to describe solid objects resistance to deformation and interaction between molecule atoms. In this model, a particle with mass $m$ is bound to a spring with elastic constant $k$, which represents the deformation resistance. If the spring is extended or compressed by an external agent, the spring will exert a force $F$ on the particle, and this force is given by the equation $F=-kx$, known as Hooke Law. The variable $x$ is the particle position and indicates extension ($x>0$) or compression ($x<0$) suffered by the spring. System stability is due to the negative signal, indicating that in response to the external influence, the force produced by the spring will exert a contrary effect.

A particle subject to Hooke Law is a system known as harmonic oscillator. Using Newton second law, a second order differential equation is obtained which its solution $x(t)$ describes the particle position in function of time.

Harmonic oscillator fractional differential equation generalization is obtained by allowing the derivative order to be a non-integer number and different than $2.0$. This small change surprisingly presents a dynamics similar to that of the damped harmonic oscillator  \cite{Achar2001}. On Physics, a damped harmonic oscillator  is obtained when a damping force, or drag, that is proportional to the negative first derivative of $x$, is added to the spring-mass system. Instead of adding a damping factor, changing the derivative order from $\alpha=2.0$ to $\alpha=1.8$ on the not damped harmonic oscillator differential equation seems enough to obtain the same behavior  of $x(t)$.

As mentioned, the harmonic oscillator is represented by a second order differential equation and is a pillar of classical mechanics. The not damped simple harmonic oscillator equation is
\begin{equation}
\frac{d^{2}x}{dt^{2}}+\omega_{0}^{2}x=0~,\label{eq:EDifOHS}
\end{equation}
where $\omega_{0}\equiv\sqrt{{k}/{m}}$ is the oscillation angular frequency. The simple harmonic oscillator solution is $x(t)$ and can be determined by mathematical methods like Laplace transform,
\begin{equation}
x(t)=Acos(\omega_0 t+\varphi)\equiv x(0)cos(\omega_{0}t)+\frac{\dot{x}(0)}{\omega_{0}}sin(\omega_{0}t)~,\label{eq:SEqOHS}
\end{equation}
where $\varphi$, $A$, $x(0)$ and $\dot{x}(0)$ are respectively the phase constant, oscillation amplitude, initial position and initial velocity.
		
On physical situations closer to macroscopic reality, mechanical energy dissipations are always present due to drag or fluid viscosity where the movement may take place. Representing the damping as a velocity proportional force $F=-\rho{dx}/{dt}$, then by Newton second law, the fractional damped harmonic oscillator differential equation becomes
\begin{equation}
\frac{d^{2}x(t)}{dt^{2}}+\gamma\frac{dx}{dt}+\omega_{0}^{2}x(t)=0~,\label{eq:EDifOHA}
\end{equation}
where we call the $\gamma=\rho/m $ parameter as damping coefficient. For the underdamped case $\gamma<2\omega_{0}$, the solution is
	\begin{equation}
	x(t)=Ae^{-(\gamma/2)t}cos(\omega t + \varphi)~,\label{eq:SEqOHA}
	\end{equation}
in which $\omega^{2}=\omega_{0}^{2}-(\gamma/2)^{2}$.	
	

\textbf{Fractional Oscillator.}

This article assumes as the definition for fractional harmonic oscillator the expression built from the not damped simple harmonic oscillator differential equation generalization, which is Eq.\ref{eq:EDifOHS}. The derivative order $2$ is changed to $\alpha$ and new variables must be used.

\begin{equation}
\frac{d^{\alpha}\xi(\tau)}{d\tau^{\alpha}}+\varpi^{2}\xi(\tau)=0~,\label{eq:EqDifOF}
\end{equation}
$\alpha$ being a real number on the interval $1.0<\alpha\leq2.0$ and $\varpi$ a constant to be defined. It is important to observe that in this equation the variables $\xi$ and $\tau$ and the parameter $\alpha$ e $\varpi$ must be dimensionless.

The Fractional Oscillator (FO) solution obtained via Laplace transform is presented by the literature\cite{Achar2001} as being the function
\begin{equation}
\xi(\tau)=\xi(0)E_{\alpha,1}(-\varpi^{2}\tau^{\alpha})+\dot{\xi}(0)\tau E_{\alpha,2}(-\varpi^{2}\tau^{\alpha})~,\label{eq:SEqOF}
\end{equation}
where $E_{\alpha,\eta}(z)$ is the Mittag-Leffler generalized function.

The Mittag-Leffler generalized function is a complex function of one complex variable and two complex parameters $\alpha$ and $\eta$ \cite{mittagFCAA}. This function was introduced in 1903 by the mathematician Gösta Mittag-Leffler and generalized in 1905 by Anders Wiman. It is defined by a power series given by:
\begin{equation}
E_{\alpha,\eta}(z)=\sum_{\begin{subarray}{c}
	k=0\\
	\\
	\end{subarray}}^{N \rightarrow + \infty}\frac{z^{k}}{\Gamma(\alpha k+\eta)}~,\label{eq:MittLeff}
\end{equation}
$\Gamma$ being the usual gamma function \cite{WolframGamma}.

\begin{figure}[ht]
	\centering{}\includegraphics[scale=0.4]{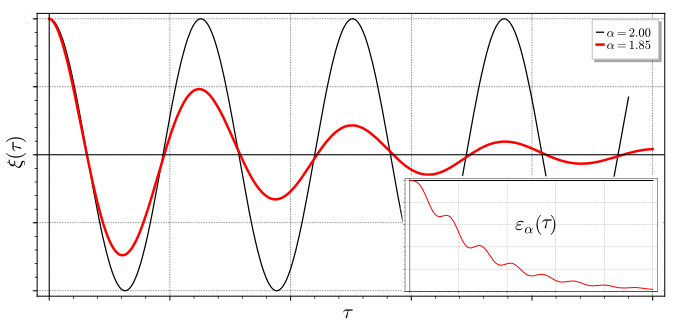}
	\caption{\label{Fig.PosicaoTempo} The bigger graph presents the function $\xi(\tau)$, Eq.\ref{eq:SEqOF}, which is the solution for the fractional oscillator differential equation with the conditions $\xi(0)=1$, $\dot{\xi}(0)=0$ and $\varpi=1$. On the inset there is the equivalent of the mechanical energy, Eq.\ref{eq:EnerOFexplicita}. On both, the physical quantities are dimensionless and on the {inset} we used $\kappa=\mu=1$. The red curve is very similar to the solution for a underdamped oscillator differential equation, but in reality, it represents the solution for a fractional oscillator differential equation. The red curve is obtained by computational implementation of the fractional oscillator solution, presented on section \ref{ResImplSoft}.}
\end{figure}

By definition, the Mittag-Leffler function has an infinite number of terms. This makes it difficult to compare it to non-elementary functions. On the next section we show that with computer algebra systems it is easy to implement the function with a finite and arbitrary number of terms, in that way, controlling the desired precision.

On this article, the Mittag-Leffler function variable and all parameters are considered as being real. Thus, it is clear that this function is a generalization of the exponential function written in it is Taylor series form
\begin{equation}
e^{\tau}=\sum_{k=0}^{\infty}\frac{\tau^{k}}{k!}=\sum_{k=0}^{\infty}\frac{\tau^{k}}{\Gamma(k+1)}=E_{1,1}(\tau)~.
\end{equation}

Just as the exponential function is the solution of linear differential equations, the Mittag-Leffler function is the solution of linear fractional differential equations with constant coefficients. The fractional oscillator solution is the function $\xi(\tau)$, Eq.\ref{eq:SEqOF}, written in terms of the Mittag-Leffler function, stressing that $\xi$ and $\tau$ are dimensionless. The graph of $\xi(\tau)$ is presented on Fig.\ref{Fig.PosicaoTempo}. This graph construction is detailed on section \ref{sec:Resultados}, it is the implementation of Mittag-Leffler function with a large number of terms on a mathematics software.

The Fig.\ref{Fig.PosicaoTempo} is a graph for Eq.\ref{eq:SEqOF} with $\dot{\xi}(0)=0$. When $\alpha=2.00$, the graph matches with non-damped simple harmonic oscillators. In fact, in this case the Mittag-Leffler functions reduces to
\begin{align}
E_{2,1}(-\varpi^{2}\tau^{2}) & =cos(\varpi \tau)~,\\
E_{2,2}(-\varpi^{2}\tau^{2}) & =\frac{\varpi}{\tau}sen(\varpi \tau)~,
\end{align}
and Eq.\ref{eq:SEqOF} becomes a solution for simple harmonic oscillators, Eq.\ref{eq:SEqOHS}.

However, for value close but less than two $(\alpha\lesssim2.0)$, a behavior very similar to that of the damped harmonic oscillator solution with low damping is observed. The possibility of a differential equation in fractional calculus representing both the simple harmonic oscillator and damped harmonic oscillator is quite interesting. The surprising fact is that damping is intrinsic to a fractional oscillator by only adjusting $\alpha$ to represent a viscosity force in a damped harmonic oscillator without the need for an extra term.

Similarly, the equivalent for a mechanical energy of fractional oscillators can be determined, the dimensionless quantity $\varepsilon$. From the literature, Ref.\cite{Achar2001}, we take the expression
\begin{align}
	\varepsilon(\tau) & =\frac{1}{2}\kappa\xi(0)^{2}[E_{\alpha,1}(-\varpi^{2}\tau^{2})]^{2}\nonumber \\
	& +\frac{1}{2}\mu\xi(0)^{2}\varpi^{4}\tau^{\alpha}[E_{\alpha,\alpha}(-\varpi^{2}\tau^{2})]^{2}~,\label{eq:EnerOFexplicita}
\end{align}
$\kappa$ and $\mu$ being dimensionless parameters corresponding to $k$ and $m$ for a damped harmonic oscillator.

In Fig.\ref{Fig.PosicaoTempo} it can be observed as mentioned that the expression $\varepsilon(\tau)$ of the fractional oscillator has a behavior similar to the mechanical energy of an underdamped harmonic oscillator.

This similarity is already known from the literature \cite{Achar2001}, and induces to question if there would be an analytical equivalence between the harmonic and fractional oscillator differential equations. In the next section the result of this work that seeks to answer this question is presented.

\section{\label{sec:Resultados}Results}

This letter involves few approximations and simplifications that make it easy to understand the issue. This section starts presenting the importance about turning variables dimensionless on fractional calculus when it is used for physical phenomena analysis. On the second part approximated analytical equivalence for the non-integer derivative order close to two is presented. Following, we show mathematical software usage, numerical regression and calculus of variations.

\subsection{\label{sec:Adimensin}Importance of nondimensionalization for the usage of fractional calculus on Physics}

Only now fractional calculus texts are being written for use in Physics area. In those texts, a connection or exemplification is sought with mathematical expressions to describe physical phenomena. Unfortunately there are divergences as to the definition of the expression for fractional oscillators and the physical meaning of the involved quantities. As an example, one version of the Eq.\ref{eq:EqDifOF} is presented by F. Mainardi in 1995 (see page 1466 from Ref.\cite{Mainardi1996}). But instead of the dimensionless variable $\varpi$, he uses the $\omega$ variable with dimension $[T]^{-1}$.

With a few expections \cite{Aguilar}, the nondimensionalization process is not observed when fractional calculus is used to analyze physical phenomena. That makes the natural phenomena interpretation more difficult. It is clear that $\xi$, $\tau$ and $\varpi$ are the dimensionless versions of physical quantities $x$, $t$ and $\omega$ for a damped harmonical oscillator. But the $\alpha$ order derivative of $\xi$ does not have the same meaning as the order $2$ derivative of $x$ does.

This confusion stems from the fact that there is no standard procedure yet for generalizing a differential equation of integer order that describes a physical phenomenon to the fractional equation equivalent. On this work, we followed four steps. First we nondimensionalized the involved physical quantities by creating new dimensionless variables. Second we rewrote the differential equation with those new variables. In the third step we generalized for the fractional derivative order and finally we looked for the solution. If desired, it is possible to rewrite the found solution in terms of the initial physical variables.


\subsection{Analytical transformation from damped harmonic oscillator to fractional oscillator: Approximation for the case $\alpha$ close to 2.}\label{sec:TransfAnalit}

On physics, to exist a direct correlation between the equation for a damped harmonic oscillator and for a fractional oscillator, Eq.\ref{eq:EqDifOF} and \ref{eq:EDifOHA} respectively, it is necessary that the parameters $\alpha$ and $\varpi$ are related to the parameters $\gamma$ and $\omega_{0}$,
\begin{equation}
\alpha=g_{1}(\gamma,\omega_{0}),\;\;\varpi=g_{2}(\gamma,\omega_{0})~.\label{eq:TransfFunc}
\end{equation}

Many authors \cite{ Herrmann, Achar2001, ALRABTAH} studied the solution for a fractional oscillator differential equation. But, at least inside the literature review that was made, no work that presented those relations, as proposed by this article, has been found.

First the position $x$ and time $t$ variables are made dimensionless on Eq.\ref{eq:EDifOHA}. It is assumed as the characteristic length, the non null initial position $x(t=0)=x_{0}$ and null initial velocity, $\dot{x}(0)=0$. As the characteristic time $T_{0}=2\pi/\omega_{0}$ is used (similar procedure can be used for the null initial position case). The dimensionless quantities $\xi=x/x_{0}$ and $\tau=t/T_{0}$ are defined. The Eq.\ref{eq:EDifOHA} after manipulation becomes

\begin{equation}
\frac{d^{2}\xi(\tau)}{d\tau^{2}}+T_{0}\gamma\frac{d\xi(\tau)}{d\tau}+4\pi^{2}\xi(\tau)=0~.\label{EqDifOFAdm}
\end{equation}
Inside the literary review, arguments that emphasized the need to nondimensionalize the quantities used on the calculus have not been found. However, for reasons of coherence with the usual calculus and the meaning of physical quantities, this step is important and necessary.

Now the Eq.\ref{EqDifOFAdm} is manipulated to show terms with fractional derivative. The Caputo definition of fractional derivative allows the next step execution,using the property given by Eq.\ref{eq:DerComutativa}. The second order derivative is split and the following is found

\begin{equation}
\frac{d^{\alpha}}{d\tau^{\alpha}}\left(\frac{d^{\beta}\xi}{d\tau^{\beta}}\right)+T_{0}\gamma\frac{d\xi}{d\tau}+4\pi^{2}\xi(\tau)=0~,\label{eq:Transf1}
\end{equation}
where $\alpha+\beta=2$. It is possible to write the fractional derivative of a function in terms of its integer derivatives \cite{Pooseh2013}. Since $1<\alpha\leq2$ then $0\leq\beta<1$ e $d^{\beta}\xi/d\tau^{\beta}$ tends to the order zero derivative for $\beta\sim0$ and tends to the first order derivative for $\beta\sim1$. Hence, the following approximation can be made
\begin{equation}
\left.\frac{d^{\beta}\xi}{d\tau^{\beta}}\right|_{0\leq\beta<1}\sim f_{1}(\beta)\xi(\tau)+f_{2}(\beta)\frac{d\xi(\tau)}{d\tau}~,\label{eq:TransAproxima}
\end{equation}
as long as $f_{1}(\beta)$ and $f_{2}(\beta)$ exists and satisfy the conditions $f_{1}(0)=1$ and $f_{2}(0)=0$. In the next section simple expressions are suggested for these functions.

Since $\xi$ and $\tau$ are dimensionless quantities, the Eq.\ref{eq:TransAproxima} has no dimensional problem, a very common difficulty present in articles that treat fractional calculus to describe physical quantities.
The Eq.\ref{eq:Transf1} becomes
\begin{equation}
f_{1}\frac{d^{\alpha}\xi}{d\tau^{\alpha}}+f_{2}\frac{d^{(1+\alpha)}\xi}{d\tau^{(1+\alpha)}}+T_{0}\gamma\frac{d\xi}{d\tau}+4\pi^{2}\xi(\tau)\sim0~.\label{eq:Transf2}
\end{equation}

Manipulating to compare with Eq.\ref{eq:EqDifOF} results in

\begin{equation}
\frac{f_{2}}{f_{1}}\frac{d^{(1+\alpha)}\xi}{d\tau^{(1+\alpha)}}+\frac{T_{0}\gamma}{f_{1}}\frac{d\xi}{d\tau}+\left[\frac{4\pi^{2}}{f_{1}}-\varpi^{2}\right]\xi(\tau)\sim0~.\label{eq:Transf3}
\end{equation}

It is desired that in the limit situation $\gamma=0$ and $\alpha=2$ ($\beta=0$ and $f_1 =1$), $\varpi=T_0\omega_0=2\pi$ is obtained. So it is defined

\begin{equation}
\varpi^{2}\equiv\frac{4\pi^{2}}{f_{1}}~.\label{DefOmegab}
\end{equation}
Hence, the term between square brackets in Eq.\ref{eq:Transf3} disappears. Rewriting Eq.\ref{eq:Transf3}

\begin{equation}
\frac{d}{d\tau}\left[\frac{f_{2}}{f_{1}}\frac{d^{\alpha}\xi}{d\tau^{\alpha}}+\frac{T_{0}\gamma}{f_{1}}\xi\right]\sim0~.\label{eq:Transf4}
\end{equation}
Now the term between square brackets must be constant. For convenience we assume it is equal to zero and by manipulating again, the following is obtained

\begin{equation}
\frac{d^{\alpha}\xi}{d\tau^{\alpha}}+\frac{T_{0}\gamma}{f_{2}}\xi=0~.\label{eq:Transf5}
\end{equation}
It is desired that Eq.\ref{eq:Transf5} is also identical to Eq.\ref{eq:EqDifOF}, that is,
\begin{equation}
\varpi^{2}=\frac{T_{0}\gamma}{f_{2}}~. \label{wGeral}
\end{equation}
and, therefore, the equality of Eq.\ref{DefOmegab} with Eq.\ref{wGeral} results in
\begin{equation}
\frac{f_{1}(\beta)}{f_{2}(\beta)}=\frac{2\pi\omega_{0}}{\gamma}~. \label{wGeral1}
\end{equation}
Thus there is a relation between $f_1(\beta)$ and $f_2(\beta)$.

Accordingly, it is possible to make an approximation that allows to transform the damped harmonic oscillator (Eq.\ref{eq:EDifOHA}) in fractional oscillator (Eq.\ref{eq:EqDifOF}). The relations between the parameters of both equations, expressed by Eq.\ref{eq:TransfFunc}, are obtained via approximated expansion. The Eq.\ref{DefOmegab} and Eq.\ref{wGeral} allow to determine $\alpha$ and $\varpi$ in funcion of $\omega_0$ and $\gamma$ for $\alpha$ close to 2. In the following we exemplify this process for one simple case.

\subsection{\label{sec:Caso simples} Simple Case: linear approximation for $f_1$ and $f_2$}

The approximation expansion given by Eq.\ref{eq:TransAproxima} depends on the functions $f_1(\beta)$ and $f_2(\beta)$. It is more accurate as $\alpha$ is inferior but very close to 2 $(\alpha\lesssim2,0)$. Therefore $\beta$ will also be positive and close to zero, $(\beta\sim0)$. When $\beta=0$, $f_{1}(0)=1$ and $f_{2}(0)=0$ must happen. Since the most simple expression is the straight line equation,

\begin{equation}
f_{1}(\beta)={1-\beta},\;\;f_{2}(\beta)={c\beta} ~. \label{eq:AproxSimples}
\end{equation}
where $c$ is an adequate constant that will be determined later.

To determine $\alpha$ and $\varpi$ we used the \ref{wGeral1} and \ref{eq:AproxSimples} equations,
\begin{equation}
\frac{1-\beta}{c\beta}=\frac{2\pi\omega_{0}}{\gamma}\rightarrow\beta = \left(1+\frac{2\pi c\omega_{0}}{\gamma}\right)^{-1}~.\label{eq:TransfFunc11}
\end{equation}
Since $\alpha = 2 - \beta$, we have the first of the relations expressed in Eq.\ref{eq:TransfFunc},
\begin{align}
\alpha & =g_{1}(\gamma,\omega_{0})=2-\left(1+\frac{2\pi c \omega_{0}}{\gamma}\right)^{-1}~.\label{eq:TransfFunc2}
\end{align}

Expanding Eq.\ref{eq:TransfFunc2} to the first order we obtain
\begin{equation}
\alpha\approx2-\frac{1}{c\pi}\left(\frac{\gamma}{2\omega_{0}}\right)~,\label{eq:TransfFunc22}
\end{equation}
under the condition
\begin{equation}
\frac{1}{c\pi}\left(\frac{\gamma}{2\omega_{0}}\right) \leq 1~. \label{eq:Condicc}
\end{equation}

In underdamped harmonic oscillators, the condition $({\gamma}/{2\omega_{0}}) < 1$ needs to be satisfied. Therefore the constant $c$, to be determined later, must satisfy $c\leq1/\pi=0.318$.

Using equations \ref{wGeral}, \ref{eq:AproxSimples}, \ref{eq:TransfFunc22}, and remembering that $\omega_0=2\pi/T_0$ and $\alpha + \beta = 2$, the following is obtained
\begin{equation}
\varpi\approx2\pi~.\label{eq:TransfFunc33}
\end{equation}

This remarkable result is valid only for the underdamped oscillation. The parameter $\gamma$ is restricted to the interval $0\leq\gamma<2\omega_{0}$. Using the Eq.\ref{eq:TransfFunc11}, this results on the interval
\begin{align}
0 & 
\lesssim\beta < \frac{1}{1+\pi c}=0.5~.\label{eq:TransfFunc3}
\end{align}
This is the interval in which the approximation Eq.\ref{eq:TransAproxima} is satisfied for the case of linear relations for $f_1$ and $f_2$ (Eq.\ref{eq:AproxSimples}).

Therefore, if the damped harmonic oscillator parameters $\omega_{0}$ and $\gamma<2\omega_0$ are known, then the fractional oscillator equation can be approximately determined. We will show below how to obtain the value for $c$ after presenting numerical results using mathematics software.

Expressions equivalent to equations \ref{eq:TransfFunc2}, \ref{eq:TransfFunc22} and \ref{eq:TransfFunc33} have not been found until the elaboration of this article.

We have not found an exact analytical procedure, without approximations, that is valid to whichever values for $\gamma$ and $\omega_0$ of the damped harmonic oscillator. This impossibility has taken us to speculate that this analytical equivalence might not be possible. This will be demonstrated in the following sections.

\subsection{\label{ResImplSoft}Software Implementation}

\begin{figure}[ht]	
	\begin{centering}
		\includegraphics[scale=0.4]{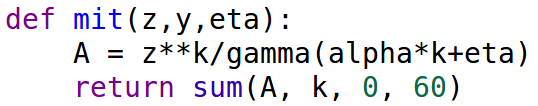}
		\par\end{centering}
	\caption{\label{Fig:CocalcMittag} Code to define the Mittag-Leffler function $E_{\alpha,\eta}(z)$ with 60 terms on the CoCalc program.}
\end{figure}

\begin{figure}[ht]
	\begin{centering}
		\includegraphics[scale=0.28]{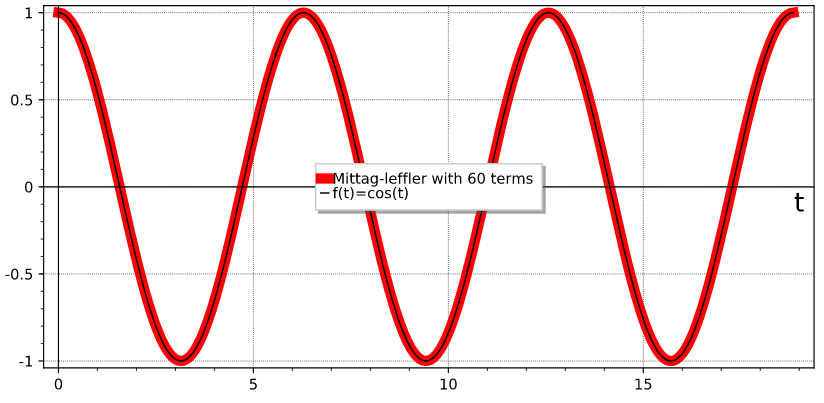}
		\par\end{centering}
	\caption{\label{Fig:CocalcMittagCoss} Mittag-Leffler Function $E_{2,1}(-t^{2})$ with 60 terms and $cos(t)$ function on Cocalc}
\end{figure}

The usage of mathematical softwares allows the use of functions defined by series with a large number of terms. For example, the Mittag-Lefler function implementation on the program CoCalc is shown on Fig.\ref{Fig:CocalcMittag}. Since it is an infinite number series, the Mittag-Leffler function is hard to be analytically manipulated. On Fig.\ref{Fig:CocalcMittagCoss} it is shown that as long as a large number of terms is used, a very good approximation is obtained between an elementary function and its correspondent representation with a Mittag-Leffler, as for example $E_{2,1}(-t^{2})=cos(t)$.

The commands contained on Fig.\ref{Fig:CocalcMittag} are simple and well documented on the manuals. It should be noted that the commands are identical or similar in different software because they are usually based on the Python language. Explaining the programming codes used is not the objective of this work, but to exemplify that advanced mathematical expressions can be easily implemented with just a few lines of code.

When manipulating this function on the program, the sufficiency of using 60 terms to represent the Mittag-Leffler function was verified. Numerical regression is shown on the next part, and how it can be used to compare two functions and quantify the approximation error. Therefore, computational numerical procedures are used as follows. This will allow to verify if there is an analytical equivalence between the fractional oscillator and the damped harmonic oscillator for all values of $\alpha$.

\subsection{\label{ResRegNum}Numerical Regression}

On Fig.\ref{Fig.PosicaoTempo} the fractional oscillator solution is shown and there is no doubt about its similarity with the damped harmonic oscillator solution. However, observing graph behaviors is not enough to confirm the equivalence between the differential equations Eq.\ref{eq:EDifOHA} and Eq.\ref{eq:EqDifOF}. For example the Gaussian function \cite{WolframGauss} and the Lorentzian \cite{WolframLorent} are graphically very similar, but they are very different solutions for different differential equations. And, more importantly , it is impossible to perfectly adjust the points generated by the Gaussian function with the Lorentzian function expression, or vice versa, using numerical regression.

The uniform and absolute convergence theorem shows that the representation of a function by a power series is unique \cite{Arfken} \cite{WolframConvUnif}. If two power series are different, they represent different functions. If they coincide, they represent the same function.

If two apparently different functions are expanded in the same interval in power series and each function expansion presents the same coefficients, then the equivalence of those functions can be confirmed. This also means that it is possible to generate points with one of the functions and use the other function to reproduce those points by adjusting the parameters. That is the idea used in this work.

The Eq.\ref{eq:SEqOF}, with the initial conditions $\xi(0)=1$ and $\dot{\xi}(0)=0$, results in 
\begin{equation}
\xi(\tau)=E_{\alpha,1}(-\varpi^{2}\tau^{\alpha})~.\label{eq:SEqOF2}
\end{equation}
This equation is implemented on CoCalc with $60$ terms. By fixating values for $\varpi$ and $\alpha$ as well as attributing values for $\tau$, the values for $\xi$ are found. The ordered pairs $\left[\tau_{i},\xi_{i}\right]$, with $i = 0, 1, 2, .., N$ play the role of experimental points.

In order to use the Eq.\ref{eq:SEqOHA} as an adjustment function of those points, first it must be nondimensionalized, resulting in 
\begin{equation}
\tilde{\xi}(\tau)=e^{-(\tilde{\gamma}/2)\tau}cos(\tilde{\omega}\tau)~,\label{eq:SEqOHA2}
\end{equation}
where $\tilde{\xi}={x}/{A}$, $\tilde{\gamma}=\gamma T_0$, $\tilde{\omega}=\omega T_0$ and $\tau=t/T_0$. Since the oscillation is underdamped, the parameter $\gamma$ is restricted to the $0<\gamma\leq2\omega_{0}$ interval. This results on the interval $0<\tilde{\gamma}\leq4\pi=12.6$.

Because of the initial conditions of Eq.\ref{eq:SEqOF2}, we also made $\varphi=0$ on Eq.\ref{eq:SEqOHA2}. This way the adjustment function is determined, $\tilde{\xi}(\tau_{i})$.\footnote{The reverse producedure, exchanging the role of the equations, is equally valid, but it is harder to execute.} Better values for $\tilde{\gamma}$ and $\tilde{\omega}$ are determined through numerical regression.

The standard error q definition is considered, given by Ref.\cite{Taylor}.
\begin{equation}
q^{2}=\frac{1}{N(N-1)}\sum_{i=0}^{N}\left[\xi_{i}-\tilde{\xi}(\tau_{i})\right]^{2}~,\label{ErroPad}
\end{equation}
where $\left[\tau_{i},\xi_{i}\right]$ is one of the $N+1$ experimental points, Eq.\ref{eq:SEqOF2}, and $\tilde{\xi}(\tau_{i})$ is the value of the adjustment function given by Eq.\ref{eq:SEqOHA2}. In a certain way, the Eq.\ref{ErroPad} is a discrete form of variational calculus \cite{WolframCalcVar}, \cite{Marion}.
\begin{equation}
q^{2}(\alpha)= \int_{\tau_0}^{\tau_{N+1}}f\left[\xi_{i}(\alpha,\tau_{i}),\tilde{\xi}(\tau_{i});\tau\right]d\tau~.\label{CalcVar}
\end{equation}

The integral Eq.\ref{CalcVar} condition to be a stationary value, in this case a minimum, is that $q^2$ is independent of $\alpha$, that is
\begin{equation}
{q^{2}}\sim \bar{q^2}\pm \delta (q^2)~, \label{EstadoEstacionario}
\end{equation}
$\bar{q^2}$ being constant and $\delta (q^2)$ the error fluctuation caused by the numerical approximation. This expressional means that the value of $q^2$ cannot depend on the value of $\alpha$. Beyond that, on the case that the equations Eq.\ref{eq:SEqOF2} and Eq.\ref{eq:SEqOHA2} are analytically equivalent, then it is also expected that  $q\sim0$ is independent on the value of $\alpha$.

\begin{table}[ht]
	\centering{}%
	\begin{tabular}{c||c|c|c}
		\hline 
		$\alpha$ & $\tilde{\gamma} = T_0 \gamma$ & $\tilde{\omega} = T_0 \omega$ & $q(10^{-6})$  \tabularnewline
		\hline
		$2.00$ & $10^{-9}$  & $6.28$ & $10^{-14}$ \tabularnewline
		\hline 
		$1.96$ & $0.39$  & $6.52$ & $0.36$ \tabularnewline
		\hline
		$1.92$ & $0.82$  & $6.77$ & $1.18$ \tabularnewline 
		\hline
		$1.90$ & $1.05$  & $6.90$ & $1.66$ \tabularnewline 
		\hline
		$1.88$ & $1.29$  & $7.03$ & $2.13$ \tabularnewline 
		\hline
		$1.84$ & $1.81$  & $7.31$ & $3.02$ \tabularnewline
		\hline
		$1.80$ & $2.39$  & $7.60$ & $3.78$ \tabularnewline
		\hline
		$1.76$ & $3.02$  & $7.90$ & $4.34$ \tabularnewline
		\hline
		$1.72$ & $3.71$  & $8.23$ & $4.66$ \tabularnewline 
		\hline
		$1.68$ & $4.46$  & $8.58$ & $4.78$ \tabularnewline 
		\hline
		$1.64$ & $5.29$  & $8.96$ & $4.79$ \tabularnewline 
		\hline
		$1.60$ & $6.23$  & $9.36$ & $4.68$ \tabularnewline
		\hline
		$1.56$ & $7.28$  & $9.79$ & $4.49$ \tabularnewline
		\hline
		$1.50$ & $9.12$  & $10.51$ & $4.06$ \tabularnewline
		\hline
	\end{tabular}
	\caption{\label{tab:Ajuste} Some of the parameters found from Eq.\ref{eq:SEqOHA2} obtained by numerical adjustment for different points created by Eq.\ref{eq:SEqOF2}, with $\varpi$ given by Eq.\ref{eq:TransfFunc33} and values of $\alpha$ between $1.50$ and $2.00$. The interval $0<\tilde{\gamma}\leq12.6$ corresponds to the underdamped harmonic oscillator regime.}
\end{table}

On Tab.\ref{tab:Ajuste}, the result from adjusting $\varpi$ according to Eq.\ref{eq:TransfFunc33} and values of $\alpha$ on interval $1.50\leq\alpha\leq2.00$ varying in $\Delta\alpha=0.02$ is found. Generally, the $\tau$ was restricted to the interval $0.0\leq\tau\leq3.0$ uniformly varying in order to grant that the quantity of points is fixated as $N+1=101$. What is important to be observed in this table is that: (i) the expression Eq.\ref{EstadoEstacionario} is not satisfied and (ii) the error value $q^2$ is only minimal for $\alpha=2.00$. The result (i) does not depend on the number of terms of Eq.\ref{eq:SEqOF2} implemented on CoCalc. If the Eq.\ref{eq:SEqOF2} and Eq.\ref{eq:SEqOHA2} equations were analytically equivalent, the error value would be approximately the same in all the lines of Tab\ref{tab:Ajuste}.

\begin{figure}[ht]
	\begin{centering}
		\includegraphics[scale=0.36]{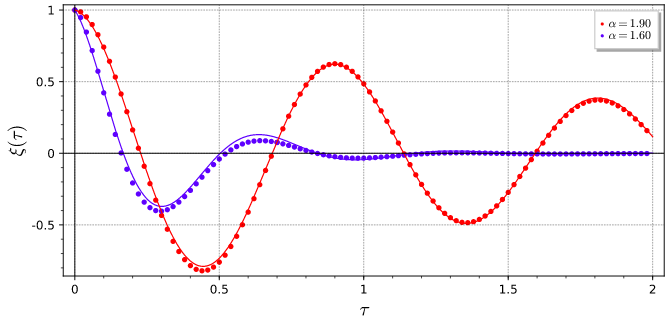}
		\par\end{centering}
	\caption{\label{fig:ResultadoOHA}Comparison of the fractional oscillator response function Eq.\ref{eq:EqDifOF}, for $\alpha=1.90$ and $\alpha=1.60$, with the damped harmonic oscillator response function Eq.\ref{eq:TransfFunc} using the parameters found by the numerical regression Tab.\ref{tab:Ajuste}.}
\end{figure}

The graph representing curves with two adjustments obtained by Tab.\ref{tab:Ajuste} is present by Fig.\ref{fig:ResultadoOHA}. It can be observed that as $\alpha$ is closer to $2.00$, the better the adjustment becomes. But as $\alpha$ moves away from this value, perfect adjustment becomes impossible.

\subsection{Test of Numerical Method}

A test was performed to verify the reliability of this method. Instead of Eq.\ref{eq:SEqOF2}, the Eq.\ref{eq:SEqOHA2} expanded with Taylor series with 60 terms in the interval  $0.00\leq\tau\leq2.00$ was used to generate the experimental points. The test results are found on Tab.\ref{tab:Ajuste2}.

\begin{table}[ht]
	
	\centering{}%
	
	\begin{tabular}{c||c|c|c}
		
		\hline 
		
		$\tilde{\gamma}_{initial}$ & $\tilde{\gamma}_{final}$ & $\tilde{\omega}_{final}$ & $q^{2}(10^{-50})$  \tabularnewline
		
		\hline
		
		$0.00$ & $10^{-9}$  & $6.28$ & $10^{-10}$ \tabularnewline
		
		\hline 
		
		$0.10$ & $0.10$  & $6.28$ & $0.14$ \tabularnewline
		
		\hline
		
		$0.20$ & $0.20$  & $6.28$ & $0.37$ \tabularnewline 
		
		\hline
		
		$0.30$ & $0.30$  & $6.28$ & $0.08$ \tabularnewline 
		
		\hline
		
		$0.40$ & $0.40$  & $6.28$ & $0.08$ \tabularnewline 
		
		\hline
		
		$0.50$ & $0.50$  & $6.28$ & $0.30$ \tabularnewline
		
		\hline
		
		$0.60$ & $0.60$  & $6.28$ & $0.24$ \tabularnewline
		
		\hline
		
		$0.70$ & $0.70$  & $6.28$ & $0.65$ \tabularnewline
		
		\hline
		
		$0.80$ & $0.80$  & $6.28$ & $1.16$ \tabularnewline 
		
		\hline
		
		$0.90$ & $0.90$  & $6.28$ & $0.87$ \tabularnewline 
		
		\hline
		
		$1.00$ & $1.00$  & $6.28$ & $0.68$ \tabularnewline 
		
		\hline
		
		$1.10$ & $1.10$  & $6.28$ & $0.11$ \tabularnewline 
		
		\hline
		
		$1.20$ & $1.20$  & $6.28$ & $0.09$ \tabularnewline
		
		\hline
		
	\end{tabular}
	
	\caption{\label{tab:Ajuste2} Some of the parameters found from Eq.\ref{eq:SEqOHA2} obtained using numerical adjustment for different points created by Eq.\ref{eq:SEqOHA2} itself, with $\tilde{\omega} = 2\pi$ and values of $\tilde{\gamma}_{initial}$ from $0.00$ to $1.25$. Observe that the error is many orders of magnitude smaller than in Tab.\ref{tab:Ajuste}.}
	
\end{table}

A similar calculus to the one on Fig.\ref{fig:ResultadoOHA} was again realized from the points of Tab.\ref{tab:Ajuste2} obtaining a linear correlation between the values of $\tilde{\gamma}_{initial}$ and $\tilde{\gamma}_{final}$ with adjustment coefficient equal to $\tilde{\gamma}_{final}$ indicating the equivalence between $\tilde{\gamma}_{initial}$ and $\tilde{\gamma}_{final}$.

Therefore the numerical regression presents that the differential equations for the fractional oscillator and damped harmonic oscillator cannot be equivalent. However an approximation can exist, such as is presented below.

\subsection{\label{sec:TransfAnalitReg}Approximated analytical transform from Damped Harmonic Oscillator to Fractional Oscillator: Constant c determination}

Using results obtained on the later subsection from the numerical regression, an approximate relation between the physical parameters $\gamma$ and $\omega_0$ of the damped harmonic oscillator and the numerical parameters $\alpha$ and $\varpi$ of the fractional oscillator can be found. Remembering that we are working in the underdamped region ${\gamma}<{2\omega_{0}}$. The functions $f_1=1-\beta$ and $f_2=c\beta$ of Eq.\ref{eq:TransAproxima} are far from being general cases, but they presented to be very appropriate.

\begin{figure}[ht]
	\begin{centering}
		\includegraphics[scale=0.3]{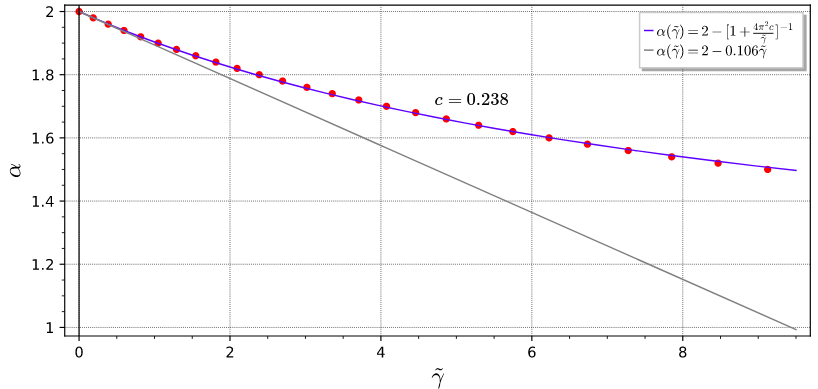}
		\par\end{centering}
		\caption{\label{fig:AlfaGamma} Relation between $\alpha$ and $\tilde{\gamma}$ according to data from Tab.\ref{tab:Ajuste}. Remembering that $\tilde{\gamma}=\gamma T_0=2\pi\gamma/\omega_0$. Regression indicates $c=0.238$.}
\end{figure}

The value of $c$ was determined using linear regression on the points from Fig.\ref{fig:AlfaGamma} obtained from Tab.\ref{tab:Ajuste}. The value $c=0.238$ was found.

Using the value $c=0.238$ on Eq.\ref{eq:TransfFunc2} yields
\begin{align}
\alpha & =g_{1}(\gamma,\omega_{0})\simeq2-\left[1+0.748\left(\frac{2\omega_{0}}{\gamma}\right)\right]^{-1}\label{eq:TransfFunc27}
\end{align}

Linearizing brings $\alpha \sim2-1.34({\gamma}/{2\omega_{0}})$, since $({\gamma}/{2\omega_{0}})<0.748$ is true. The other fractional oscillator parameter was already determined by Eq.\ref{eq:TransfFunc33}
\begin{equation}
\varpi=g_{2}(\gamma,\omega_{0})\approx 2\pi\label{eq:TransfFunc37}
\end{equation}

The value of $c=0.238<0.318$ allows us to recalculate the interval of $\alpha=2-\beta$ from Eq.\ref{eq:TransfFunc3}:

\begin{align}
1.43 \lesssim \alpha \lesssim 2.00~. \label{eq:FaixaAlfa}
\end{align}

This is the interval of validity to the approximations represented by equations \ref{eq:TransAproxima} and \ref{eq:AproxSimples}. The equations \ref{eq:TransfFunc27} and \ref{eq:TransfFunc37} are the answers sought by the expression in Eq. \ref{eq:TransfFunc}.

Hence, the analysis realized by numerical regression has shown the exact analytical equivalence between the fractional oscillator and damped harmonic oscillator to be impossible to exist. Yet, the data obtained from the regression allows, from the physical parameters $\gamma$ and $\omega_0$ of the damped harmonic oscillator, the determination in an approximate manner of the numerical parameters $\alpha$ and $\varpi$ of the fractional oscillator. This approximation is valid for the underdamped harmonic oscillator regime ${\gamma}/{2\omega_{0}}<1$.

\section{Conclusions}

The results of this work show that the functions indicated by equations \ref{eq:SEqOHA} and \ref{eq:SEqOF} are not analytically equivalent. They can be similar when presented in graphical form, but this similarity does not correspond to analytical equivalence, in other words, perfect interpolation. Hence, their respective differential equations also do not have equivalence. That means an exact analytical transformation  for the two differential equations, fractional oscillator and damped harmonic oscillator, is impossible for a fractional derivative order of $\alpha$ different from an integer number.

However, an approximate relation between their respective parameters can be observed, in such a way that is is possible to correlate them. For example, inside the value range presented, with the damped harmonic oscillator parameters it is possible to build a fractional oscillator with a similar behavior, resulting in an approximation for the equations Eq.\ref{eq:SEqOHA} and Eq.\ref{eq:SEqOF}

In this work the computer algebra system importance nowadays to analyze physical phenomena and their differential equations was presented. Fractional calculus concepts applied to a mass-spring system were presented in an easy to comprehend language for physics students. The importance of nondimensionalization when analyzing fractional calculus usage on Physics  was highlighted. An approximate analytical transformation from a damped harmonic oscillator to a fractional oscillator was realized. For the underdamped case a relation between classical and fractional parameters was found, on the simple case approximation. For the general case, in a very easy way the Mittag-Leffler function was implemented and the numerical regression on CoCalc was made, and using variational calculus concepts, it has been demonstrated that there is no analytical equivalence between the fractional and damped oscillators.

Nevertheless, as was observed from the results, there is an approximate correlation between the parameters of the two oscillators so that the response functions of both are approximate. For the general case, of the regression results, an analytical approximation was adjusted and new relations between the oscillators parameters were obtained, that beyond being able to explain the found results, adjust very well to the fractional order derivative, for example, to the exponential-cosine kind of functions.

Finally, it is reasonable to assume there could exist physical phenomena that are better described by the fractional oscillator than by the damped harmonic oscillator, since the oscillation behavior with decay is very diversified. Future works can study the generalities of this approximation in deeper details, that can contribute to a standard procedure construction for the generalization of a differential equation order.

This result was obtained without use of sophisticated analytical mathematics. Instead, modern mathematical software resources weere utilized, the use of power series with many terms, numerical regression and non-elementary functions construction with  programming language.
 
\bibliography{article}

\end{document}